\documentclass[12pt,preprint]{aastex}

\shorttitle{$H_0$ =55}
\shortauthors{Arp}

\begin{document}

\title{Arguments for a Hubble Constant near $H_0$ = 55}

\author{H. Arp}
\affil{Max-Planck-Institut f\"ur Astrophysik, Garching 85741, Germany}
\email{arp@mpa-garching.mpg.de}

\begin{abstract}
Recent analyses of Cepheid distances to spiral galaxies have led to an announcement of a Hubble 
constant of $H_0 = 72 \pm 8$ km/sec/Mpc. The new Cepheid distances, however, show that there
are numerous redshift distances with large excesses which cannot be due to peculiar velocities. 
Ignoring these discordant redshifts, if the Hubble constant is calibrated with Cepheid distances of low 
redshift spirals a value near $H_0$ = 55 is obtained. Use of independent
distance criteria such as Tully-Fisher and group membership verifies this value and leads to three
conclusions: 1) The peculiar velocities of galaxies in space are characteristically small. 
2) Sc companions to normal Sb's tend to be less luminous, with younger stellar populations and
small amounts non velocity redshift. 3) ScI and other purportedly over luminous spirals have large
amounts of intrinsic redshift.
\end{abstract}



\section{Introduction}

Ever since Hubble's discovery of the redshift-apparent magnitude relation for galaxies, the value of the
slope of the inferred redshift-distance law has been in contention. In the 1960's there were two camps,
one claiming a value around 50 and the other claiming around $H_0$ = 100 km/sec/Mpc. The latter, high
figure in recent years has come down to values in the 70's. Now, with the results from the Hubble 
space Telescope measurement of distances to galaxies with Cepheids, the best 
value has been announced to be $H_0 = 72 \pm 8$ (Freedman et al. 2001).

It is interesting to note that in the Virgo Cluster the spirals always gave higher $H_0$'s because their 
redshifts were on average higher. It was argued by G. de Vaucouleurs and others that this was
because there was a group of more distant spirals just in back of the Virgo Cluster. This proposed
segregation has been put finally to rest with the new HST 

\begin{table}[t]
\caption{Galaxy Distances\label{tab:galdist}}
\vspace{0.4cm}
\begin{center}
\begin{tabular}{|c|c|c|c|l}
\hline
& & & & \\
Galaxy  ~~  Type &
$Cepheids (D_Z)$ &
$  H_0 = 72$ &
$  H_0 = 55$ &  
\\ \hline

NGC1365 ~SBbI &
17.95 Mpc &
21.7 Mpc &
28.4 Mpc & \\

NGC4321~ ScI~~ &
15.21~~~~~~~~&
20.3 ~~~~~~~&
26.6 ~~~~~~~& \\

NGC4535~ SBcI~&
15.78~~~~~~~~&
25.3 ~~~~~~~&
33.1 ~~~~~~~& \\

NGC4536~ ScI~~ &
14.93~~~~~~~~&
22.9 ~~~~~~~&
29.9 ~~~~~~~& \\

& & &  \\ \hline
\end{tabular}
\end{center}
\end{table}

Cepheid distances which show these high
redshift spirals to be firmly at the distance of the center of the Cluster.  (See for e.g. Table 1 of the
present paper.) Their high redshift must now be explained by something else. Peculiar velocities will
not work because the redshifts are systematically positive. For field galaxies peculiar velocities of the 
order of 1000 km/sec are out of the question because they would blow apart the whole lower third of 
the Hubble redshift-apparent magnitude relation. 

The long debated and ever changing value of the Hubble constant certainly involves excess 
redshifts, particularly of the spirals. It is the purpose of the following paper to argue that the new 
Cepheid measures actually require a value of $H_0$ only slightly greater than 50. One consequence is 
that spiral galaxies can contain a component of non-velocity redshift. Independent 
supporting evidence for this result is reviewed.

\section{The Cepheid Evidence}

Table 1 lists the final adopted distances for the luminosity class I Galaxies in Table 4 of Freedman et al.
2001. (Corrected for metallicity and called $D_z$ in the above reference.) In the last two columns of 
Table 1 above are listed the redshift distances for these galaxies, One column for $H_0$
= 72 and the last column for $H_0$ = 55. Since the Cepheid distances are the most accurate direct
measure of distance available they force the conclusion that the redshift distances for these particular
galaxies are inescapably too large.

If one uses the $v_0$ redshifts from Sandage and Tamman (1981) and divides by the Cepheid
distances listed in Table 1, one gets $H_0$ = 87, 96 ,115, and 110. Because the Cepheid distances are
so accurate it can be finally established that the Hubble constant discrepancies lie in the redshifts not 
the distances.

Are these excess redshifts due to peculiar velocities? If so why would they be all in the same
direction and 
why most conspicuously in the SI's? Actually this systematic redshift excess has been 
demonstrated previously in analyses using independent distance indicators. 

\section{Redshift Distances Compared to Tully-Fisher Distances}

As is well known, Tully-Fisher distances rely on a mass calculated from rotation which is converted to
luminosity by means of a mass-luminosity relation and hence into a distance modulus. Calibrated with
nearby galaxies whose distances are best known, they represent, in principle,  distance determinations
independent of the redshift of the galaxy.

Fig. 1 shows the differences in distances derived from redshifts relative to the distances derived from
Tully-Fisher measures. The $d_z - d_{TF}$ is plotted against blue luminosity from an analysis 
published by Arp (1990) using $H_0$ = 65. From a best fit of Cepheid distances D. Russell (2001)
reports an $H_0$ = 60. But the
deviations for the most luminous galaxies would be even larger using $H_0$ = 55 as advocated in the 
present paper. The diagram illustrates that for low redshift Sc galaxies the T-F distances agree 
well with the redshift distances. But for larger redshift galaxies the redshift distances are increasingly
in excess, reaching the rather staggering discrepancy of up to 30 Mpc! {\it It is among the highest
distance discrepancies that we find the ScI galaxies, the same highest luminosity class which gave 
such excessive Hubble constants in Table 1 using Cepheid distances.}

It is now important to plot a diagram similar to Fig. 1 using the new HST Cepheid distances.

\section{Redshift Distances Compared to HST Cepheid Distances}

Fig. 2 shows the redshift differences between distances derived using an $H_0$ = 55 and 
distances announced for the Cepheids in Freedman et al. (2001), i.e.  $d_z(55)$ - $D_z(Ceph)$.  The 
result is that the highest redshift Sc galaxies show the greatest excess compared to Cepheid
distances just as they did when compared to Tully-Fisher distances in Fig. 1. Also, as in Fig. 1, the low
redshift Sc's show good agreement with their redshift distances.

Fig. 3 now shows the same plot for Sb galaxies. (Points from a new Cepheid distance from Macri
et al. 2001 and Saha et al. 2001 have been added.) {\it The result is that the high redshift Sb's show 
the same increasing discrepancy toward higher redshift.} 

Other important properties that Figs. 2 and 3 show: 

\begin{figure}[p]
\plotone{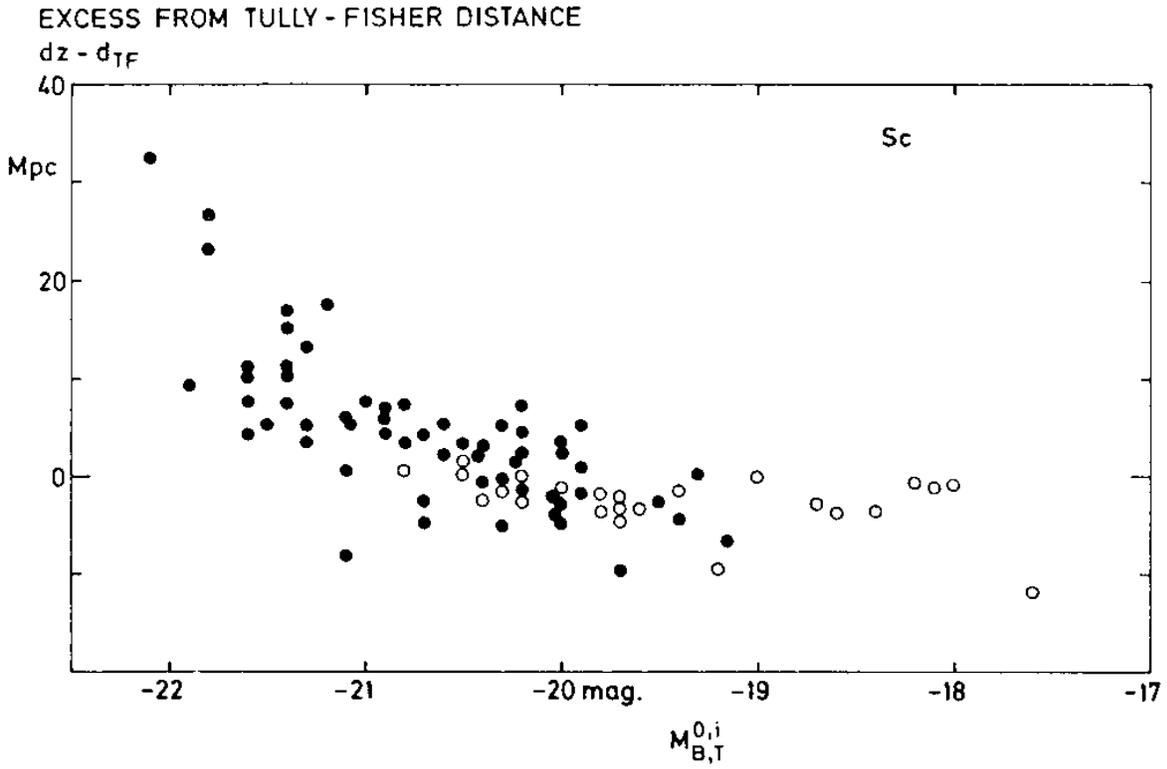}
\caption{The excess of redshift distance over Tully-Fisher distance ($d_z - d_{TF}$) 
is plotted as a function of the blue luminosity as derived in  Arp (1990). The open circles are for Sc's
with $v_0 < 1000 km/sec$ and demonstrate that redshift and Tully=Fisher distances agree well
for low redshift galaxies over a wide range in luminosity.
\label{fig1}}
\end{figure}

\begin{figure}[p]
\plotone{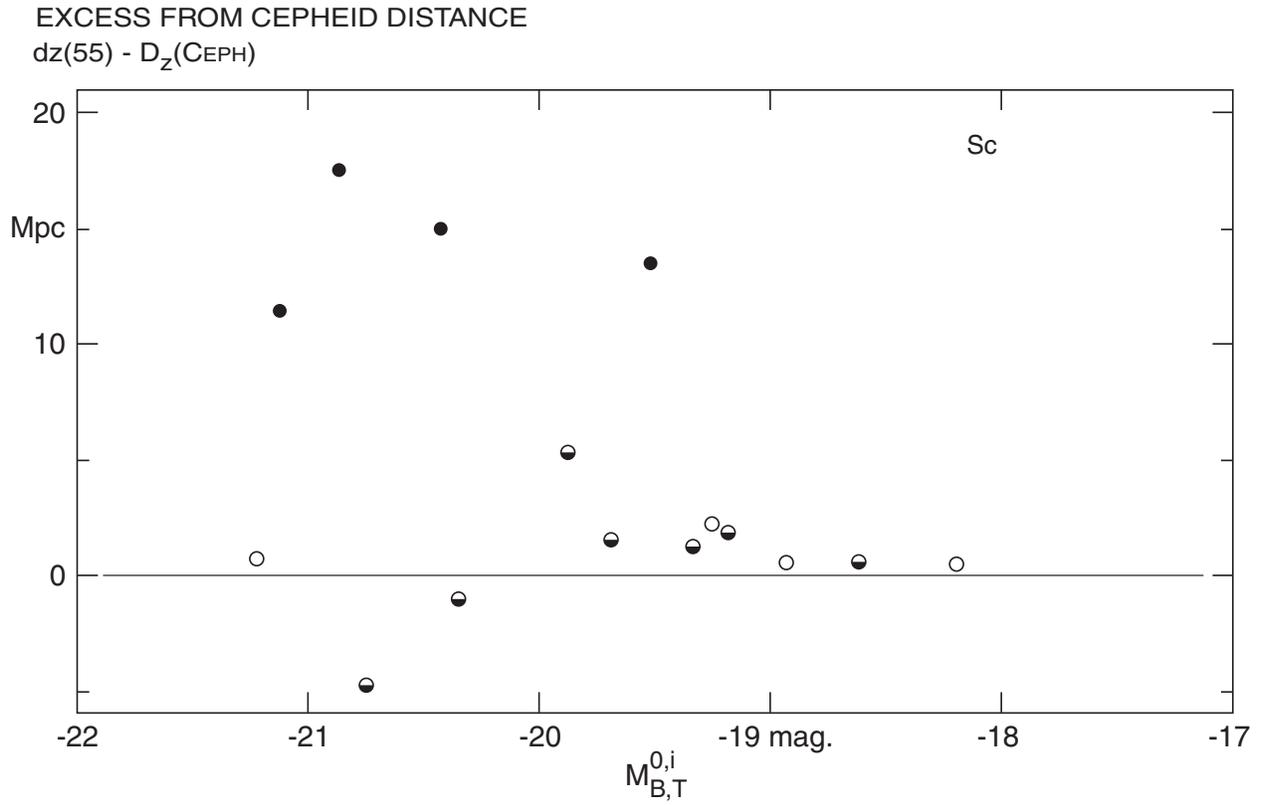}
\caption{Differences between distances using $H_0$ = 55 and distances from for the Cepheids in 
Freedman et al. (2001), i.e.,  $d_z(55)$ - $D_z(Ceph)$. Open circles represent galaxies with redshifts 
$69 \leq v_0 \leq 372$, the half filled circles $435 \leq v_0 \leq 792$ and the filled circles 
$1464 \leq v_0 \leq 1818$ km/sec. Absolute mags. are at Cepheid distances.
\label{fig2}}
\end{figure}

\begin{figure}[p]
\plotone{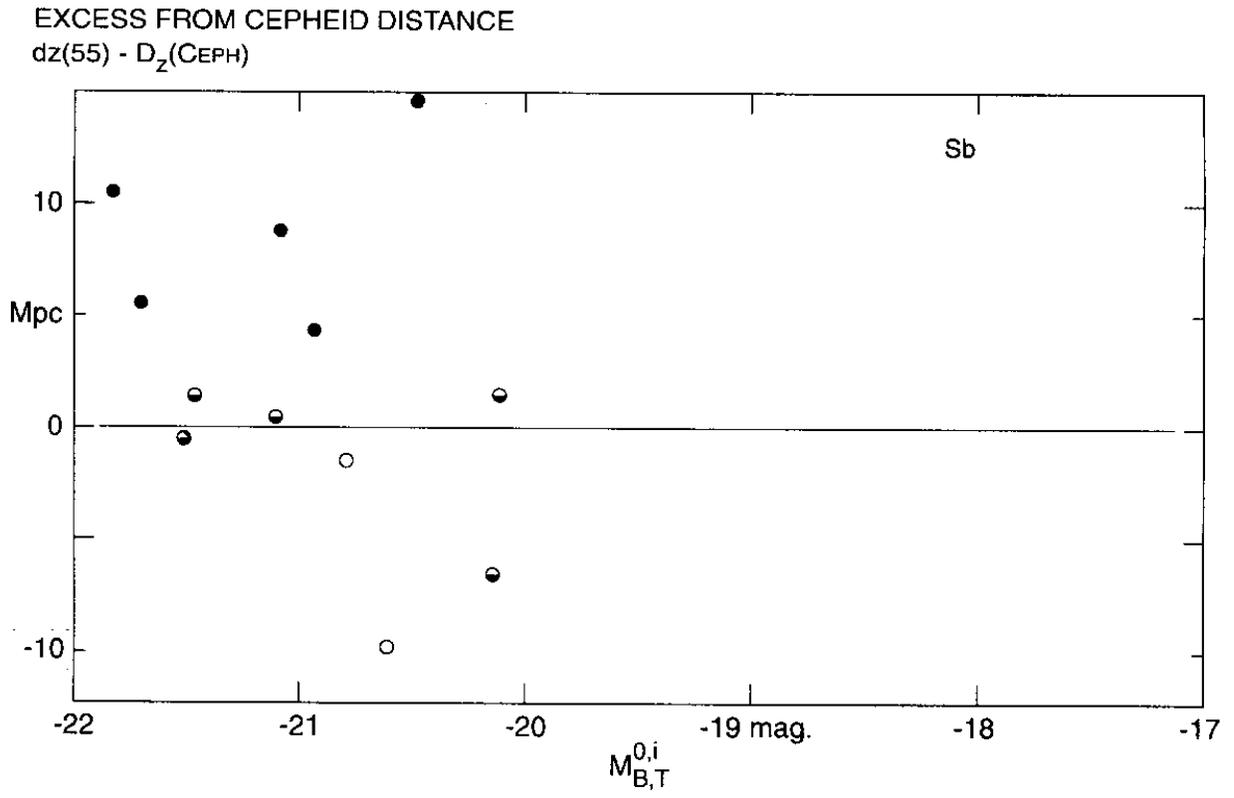}
\caption{The same excess of redshift distance from Cepheid distance is plotted for Sb galaxies. The
open circles are $0 \leq v_0 \leq 366$, half filled circles $366 < v_0 \leq 860$ and filled circles 
 $1114 \leq v_0 \leq 1577$ km/sec.
\label{fig3}}
\end{figure}

1) The Sb's are, as class,  brighter than $M_B$ = -20 mag. The Sc spirals are generally fainter than
$M_B$ = -20. This is in the sense that Sc's are usually found as fainter companions to a dominant
Sb in a group (like M31 and M81 group, see Arp 1990; 1994; 1998a,b). 

2) In the above references fainter companion Sc's are systematically redshifted from the redshift of 
the central Sb. This same effect is seen in Fig. 2 where the low luminosity Sc's are shifted slightly 
above the zero difference line. In Arp 1990, Table Ib, some of the most certain Sc companions of Sb's 
are listed with
excess redshifts of from +21 to + 599 km/sec with an average of about +174. The average upward
displacement of the Sc's fainter than $M_B$ = -20 mag. in Fig. 2 is about 1.5 Mpc or about a 
systematic redshift excess of around 83 km/sec. 

3) It might be better to try to set the value of $H_0$ with Sb spirals. But we see from Fig.
3, that {\it the same type of galaxies, with their only difference being their high redshift, give much too 
great redshift distances compared to their real (Cepheid) distances.}

\subsection{Peculiar Velocities}

It is clear from both Fig. 1 and Fig. 2 that the high luminosity Sc's and Sb's have unacceptably large
redshifts for their distances. In the past such deviations have been attributed to peculiar
(non-Hubbble flow) velocities. There are, however, four outstanding reasons why this explanation 
cannot be invoked here:

1) The spirals with low redshift that we know to be nearby have very small peculiar velocities. We
can see from Figs. 1 and 2 particularly that more than a few hundred km/sec would disrupt the good
agreement between their Cepheid and redshift distances. There seems to be no obvious reason why
locally galaxies should be anomalously quiet.

2) The largest excess redshifts as seen in Figs. 2 and 3 are essentially all positive (nine out of nine). 
If these were peculiar velocities we would expect as many negative as positive discrepancies.

3) The largest excess redshifts are from galaxies with recognizably different morphology, namely
predominantly luminosity class I galaxies. There is no reason for just these kinds of galaxies to be
expanding faster than the rest.

4) It has been argued in the past by Sandage and Tamman (e. g. 1981) that the Hubble flow is quiet to 
$\leq$ 50 km/sec. Even smaller peculiar velocities are required not to wash out redshift periodicities of
72 km/sec (Tifft 1976; Arp and Sulentic 1985) and 37.5 km/sec (Napier and Guthrie 1993).

Of course we have direct evidence from clusters and groups of galaxies that the Sc spirals and lower
luminosity companions in general have non-velocity redshifts. In the two nearest groups centered on
Sb's, M31 and M81, 22 out of 22 major companions have positive redshifts with respect to the central
galaxy (Arp 1994). More distant groups show the same pattern. Then there are the cases where high 
redshift ScI's are actually linked by luminous connections to much lower redshift galaxies 
(Arp 1990).

\section{What is $H_0$?}

Fig. 2 shows that $H_0$ = 55 fits reasonably well the lower luminosity Sc's if they have
approximately 100 km/sec intrinsic redshift. That value is loosely corroborated by the low luminosity 
Sb's in Fig. 3 but the accuracy is obviously low. The question arises, can we support this
value with a different set of galaxies? The answer is that we can by using the
accepted members of the Virgo Cluster of classes: E + E/S0, S0 + SB0 + S0p and Sa + Sab. These 55
galaxies, give a luminosity weighted mean redshift of $v_0$ = 863 km/sec (Arp 1988a). This weighting 
does not allow faint, small mass objects to overly affect the final average. The
derived value represents {\it the redshift of the average mass} residing in the major galaxies in 
the center of Virgo.

Taking the Cepheid distance to Virgo as 15.3 Mpc (from Freedman et al 2001, Table 8) we obtain a 
Hubble constant of $$H_0 = 863/15.3 = 56.4 km/sec/Mpc$$ Alternatively we could take the redshift 
of the brightest galaxy in Virgo, the central E/S0, M49, and obtain $$H_0 = 822/15.3 = 53.7
km/sec/Mpc$$

It is enlightening to notice that in Table 2 of Arp 1988 the luminosity weighted mean redshift of the 
Sb +SBb class galaxies is 97 km/sec and the same mean for the Sc + Sbc classes is 1344 km/sec.
These would give wildly implausible Hubble constants and yet it is these very classes which the
Freedman et al. (2001) Cepheid calculations of $H_0$ are based - as per Figs. 2 and 3.

The most illuminating diagram for understanding why there is a long history of contention over
the value of the Hubble constant is shown here in Fig. 4. It is clear that the redshift independent
distances to the nearest galaxies define a Hubble relation just slightly greater than $H_0$ = 50. But as
one goes past about $d_{TF}$ = 15 Mpc one encounters galaxies which have excess redshifts. Most of
these are high luminosity or ScI spirals. 

\begin{figure}
\plotone{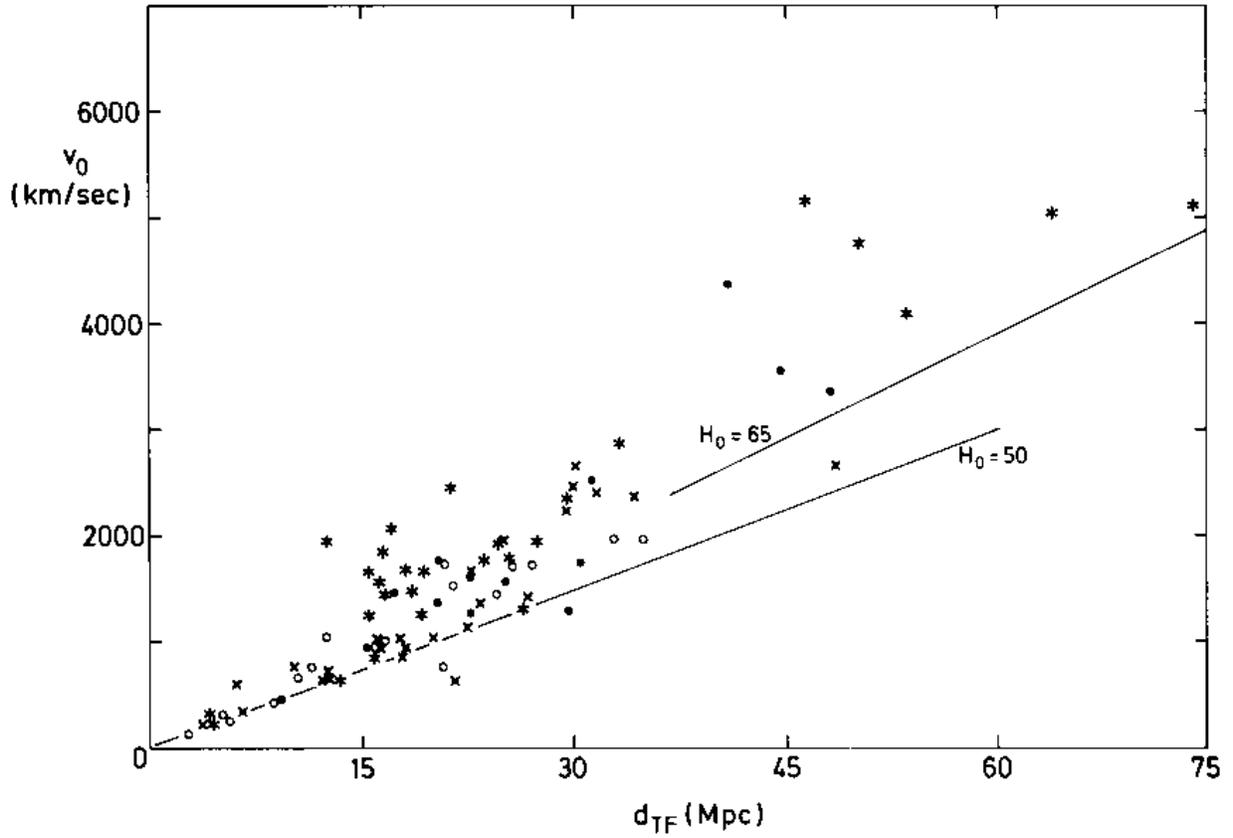}
\caption{The Tully-Fisher distance ($d_TF$) is plotted against Local Group centered redshift
($v_0$) for measured Sc's. For low redshift galaxies a very accurate fit to $H_0$ = 55 is evident. For
higher redshift galaxies Hubble constants of $H_0 \geq 100$ must result. Data and Figure from 
Arp (1990).
\label{fig4}}
\end{figure}

\begin{figure}
\plotone{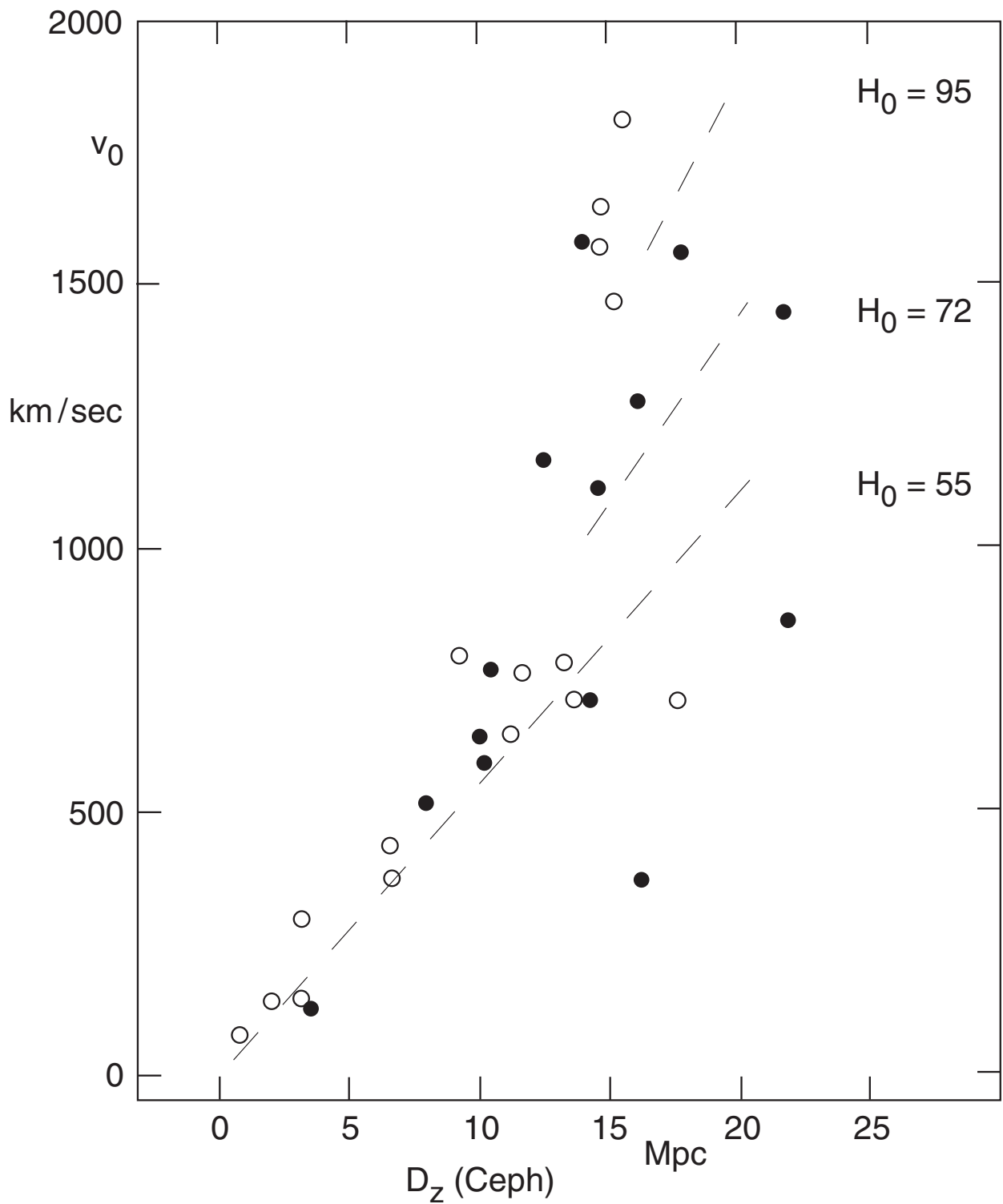}
\caption{The Cepheid distance, $D_z$(Ceph),  is plotted against Local Group centered redshift
($v_0$) for available galaxies. For low redshift galaxies a very accurate fit to $H_0$ = 55 is evident. 
Past 15 Mpc, however, excess redshifts appear which are too large and too positive to be peculiar
velocities. Filled circles are Sb, open circles Sc.
\label{fig5}}
\end{figure}

What has happened in the controversy is that in an attempt to
get further out in distance in order to minimize the effect of presumed peculiar velocities,
investigators have taken more and more galaxies with excess (intrinsic) redshifts. Naturally the
Hubble constant increases - and also becomes very indeterminate. On the other hand, previous
analyses which obtained $H_0$ near 55 (e.g. Sandage 2000; Theaureau et al. 1999) used distance 
limited samples to avoid Malmquist bias. But the "long distance scale" was involved and the sample 
was redshift-distance limited which excluded galaxies with large intrinsic redshifts.

The Tully-Fisher plot in Fig. 4 is strikingly confirmed by using Cepheid distances in Fig. 5. In the latter 
figure we can be sure the more luminous spirals are at their plotted distance and that their redshifts 
indicate too high an expansion velocity for that distance. Arguments against these discrepancies being
peculiar velocity are given in section 4.1.

\section{Remarks on some Discordant Points in Figs. 2 and 3}

In Fig. 3 both strongly negative Sb points come from the Virgo Cluster and one has an obviously low
redshift of $v_0$ = 366 km/sec. The latter is a good candidate for an older galaxy if the intrinsic
redshifts are a function of the age of their constituent matter (Arp 1998b). 

In Fig. 2 the discordant point at $M_B$ = -19.52 mag. and excess redshift distance of 13.65 Mpc is 
NGC 4496A. This is an interesting object because there is a smaller galaxy with cz = 4510 km/sec 
only 0.87 arcmin from A which has cz = 1725 km/sec. What is going on? The solution I would suggest
is that NGC 4496A was a high redshift ScI like the points to its left. But the interaction with the high
redshift companion (the interaction is argued in Arp 1990b) destroyed the sharp star forming arms
characteristic of the ScI, reduced its luminosity in the process, and turned it into the SBcIII-IV which
it presently classified. Note that a galaxy of this morphological type is ordinarily a low luminosity, 
dwarf type object.

The most difficult point in Fig. 2 is the open circle at $M_B$ = -21.24 mag. which represents the
nearby ScI galaxy M101. Its placement in the sky and redshift would argue for its being a companion
to M81. That would make its luminosity and absolute diameter much more in consonance with other
galaxies of its type. However, its Cepheid distance would have to be reduced from 6.7  to around 3.6
Mpc. As it stands it is an anomalous point in Fig. 2. Perhaps its cepheids are very low metal content
and are vibrating in a low luminosity/ low overtone mode.

\section*{References}

\noindent Arp, H. 1988a, A\&A 201, 70
\medskip

\noindent Arp, H. 1988b, Seeing Red: Redshifts, Cosmology and Academic Science, Apeiron, Montreal
\medskip

\noindent Arp, H. 1990a, Astro. Phys. Sp. Sci. 167,183
\medskip

\noindent Arp, H. 1990b, P.A.S.P. 102, 436
\medskip

\noindent Arp, H. 1994, ApJ 430, 74
\medskip

\noindent Arp, H. 1998, ApJ 496, 661
\medskip

\noindent Arp, H., Sulentic, J. 1985, ApJ 291, 88
\medskip

\noindent Freedman,W., Madore, B., Gibson, B. et al. 2001, ApJ 553, 47
\medskip

\noindent Macri, L., Stetson, P., Bothun, G., et al. 2001, ApJ in press and astro-ph/0105491
\medskip

\noindent Napier, W., Guthrie, B. 1993, Progress in New Cosmologies, Plenum Press, 29.
\medskip

\noindent Russell, D. 2001, ApJ submitted
\medskip

\noindent Saha, A., Sandage A., Thim, F. et al. 2001, ApJ 551, 973
\medskip
 
\noindent Sandage, A. 2000, PASP 112, 504
\medskip

\noindent Sandage, A., Tamman, G. 1981, A Revised Shapley-Ames Catalog of Bright Galaxies,
Carnegie Institution of Washington 
\medskip

\noindent Theureau, G., Hanski, M., Ekholm, T., Bottinelli, L., Gougenheim, L., Pateurel, G., Teerikorpi,
P. 1999, IAU 183, 71
\medskip 

\noindent Tifft, W. 1976, Apj 206, 38
\medskip

\end{document}